\documentclass[aps,pra,twocolumn,superscriptaddress]{revtex4}
\usepackage{amsfonts}
\usepackage{amsthm,amsmath,amssymb}
\usepackage{mathrsfs}
\usepackage{epsfig}
\usepackage{color}
\usepackage{graphics, graphicx}
\usepackage{bbold}
\usepackage{psfrag}
\usepackage{mathcomp}
\usepackage{subfigure}
\usepackage{verbatim}
\usepackage{color}
\usepackage[colorlinks,citecolor=blue]{hyperref}
\usepackage{csquotes}
\usepackage[utf8]{inputenc}
\usepackage{amsfonts}  

\begin{document}
	
	\title{Spin-imbalanced fermion on a dynamic lattice}
	
	\author{Jie Liu}
	\affiliation{State Key Laboratory of Quantum Optics Technologies and Devices, Institute
		of Laser spectroscopy, Shanxi University, Taiyuan 030006, China}
	\affiliation{Collaborative Innovation Center of Extreme Optics, Shanxi University,
		Taiyuan, Shanxi 030006, China}
	\author{Xiaofan Zhou}
	\email{zhouxiaofan@sxu.edu.cn}
	\affiliation{State Key Laboratory of Quantum Optics Technologies and Devices, Institute
		of Laser spectroscopy, Shanxi University, Taiyuan 030006, China}
	\affiliation{Collaborative Innovation Center of Extreme Optics, Shanxi University,
		Taiyuan, Shanxi 030006, China}
	\author{Suotang Jia}
	\affiliation{State Key Laboratory of Quantum Optics Technologies and Devices, Institute
		of Laser spectroscopy, Shanxi University, Taiyuan 030006, China}
	\affiliation{Collaborative Innovation Center of Extreme Optics, Shanxi University,
		Taiyuan, Shanxi 030006, China}
	
	\begin{abstract}
		We investigate the magnetic order of a one-dimensional spin-1/2 fermion dynamical lattice, where itinerant fermions are coupled to bond-centered localized spins via an Ising-like spin dependent hopping. The model provides an anisotropic dynamical extension of conventional spin-1/2 fermion systems, in which the motion of itinerant fermions is directly modulated by the configuration of localized spins. Using density matrix renormalization group simulations, we map out the ground state phase diagram in various parameter spaces. Depending on the interplay among the hopping dependent on localized spins, the longitudinal field, and the external Zeeman field, two distinct phases are obtained: a paramagnetic phase and a spin-density-wave phase. Most notably, in the partially spin-polarized fermion phase, the spin-density wave ordering wave vector exhibits two distinct phenomena, corresponding respectively to the nesting vectors $2k_{F\uparrow}$ and $2k_{F\downarrow}$ of the spin-resolved Fermi surfaces. We further demonstrate that the two spin-density wave phases are robust against the repulsive Hubbard interaction between itinerant fermions. Our results reveal a novel route for tuning magnetic modulations in one-dimensional correlated systems and enrich the microscopic understanding of dynamical lattice magnetism.
	\end{abstract}
	
	\maketitle
	
	
	\section{INTRODUCTION}
	
	Magnetism represents one of the most fundamental and challenging topics in condensed matter physics~\cite{1,2,3,4}. From conventional ferromagnetic and antiferromagnetic orders to exotic magnetic states emerging in strongly correlated quantum systems~\cite{5,6,7,8,9}, magnetic phenomena not only reflect spontaneous symmetry breaking at the macroscopic level, but are also intimately connected to frontier discoveries such as high-$T_c$ superconductivity, quantum criticality, and topological phases of matter~\cite{10,11,12,13}. In low-dimensional systems, strong quantum fluctuations and enhanced correlation effects give rise to magnetic behaviors that fundamentally differ from those in higher dimensions~\cite{DagottoRice1996}. As a result, one-dimensional (1D) magnetic systems serve as an ideal platform to explore unconventional quantum phases and correlated metallic states~\cite{14,16,17,super1,super2}.
	
	In 1D strongly correlated systems, such as the extended Fermi–Hubbard model~\cite{18,23,24,jiexi,15} and the Ising–Kondo lattice model~\cite{17,21,22}, the spin-density wave (SDW) constitutes one of the most common forms of long-range magnetic ordering~\cite{25,26}. Its presence is typically identified through the characteristic peak structure in the localized spin structure factor $S(k)$, which serves as a key diagnostic of magnetic modulation~\cite{15,23,24}. Early theoretical studies established that, in Tomonaga-Luttinger liquids and the 1D extended Fermi–Hubbard model, the dominant SDW ordering vector follows the well known relation $k=2k_F$, where for a two-component balanced fermion system $k_F = \pi \rho / 2$ with total filling density $\rho = N / L$~\cite{27,28,29}. This momentum-density relation has been rigorously verified by analytical solutions and Luttinger liquid theory~\cite{15,18,23,24,jiexi}, and further supported by large-scale numerical simulations including density matrix renormalization group (DMRG) and quantum monte carlo (QMC) studies~\cite{20,41,qmc1,qmc2}. These results establish a solid theoretical and computational foundation for understanding SDW ordering in two-component balanced 1D fermion systems.
	
	So far, previous studies have established a solid foundation for fermion spin balanced systems, introducing a spin imbalance between the two fermion components $(N_{\uparrow}\neq N_{\downarrow})$ adds a new degree of freedom in low-dimensional systems, giving rise to a rich variety of physical phenomena~\cite{32,33,39}. It can stabilize a variety of unconventional quantum phases, including the Fulde–Ferrell–Larkin–Ovchinnikov (FFLO) state, breached-pair (BP) phases, partially polarized metallic states, and phase-separated ferromagnetic regions~\cite{35,37,38,39,40}. Extensions of the Fermi–Hubbard model under fermion spin imbalance have demonstrated the emergence of additional SDW peaks controlled by the minority-spin density, coexisting with FFLO pairing oscillations~\cite{41,42}. Despite these advances, most studies have focused on pairing correlations and finite-momentum superconductivity, while the magnetic properties, particularly the evolution of the dominant SDW peak and its ordering wave vector under fermion spin imbalance in dynamical lattice systems, remain largely unexplored.
	
	In this paper, we perform DMRG simulations to investigate the ground state magnetic properties of a 1D spin-1/2 fermion dynamical lattice model~\cite{dmrg1,dmrg2,ITensor}. We mainly focus on how fermion spin imbalance affects the dominant SDW peak in the localized spin structure factor.Our numerical results reveal two distinct phenomena characterizing the SDW ordering wave vector in the fermion spin partially polarized phase.  Transitions between these two phenomena can be driven by tuning the coupling parameters $\alpha$, $\Delta$, and $\rho$. These results differ fundamentally from the conventional expectation $k=2k_F$ in single or two-component balanced fermion systems, highlighting the crucial role of fermion spin imbalance in shaping SDW modulation. Moreover, we show that manipulating the coupling parameters enables controllable transitions among SDW and paramagnetic (PM) phases in the phase diagram. We further examine the effect of repulsive Hubbard interactions, and find that the repulsive Hubbard interaction enhances the fermion spin polarized phase by strengthening the effective exchange field, while the two distinct SDW phenomena associated with partially polarized fermions remain robust within their respective parameter regions.
	
	\section{MODEL AND METHOD}
	\label{Model and Hamiltonian}
\begin{figure}[t]
		\centering
		\includegraphics[width=0.5\textwidth]{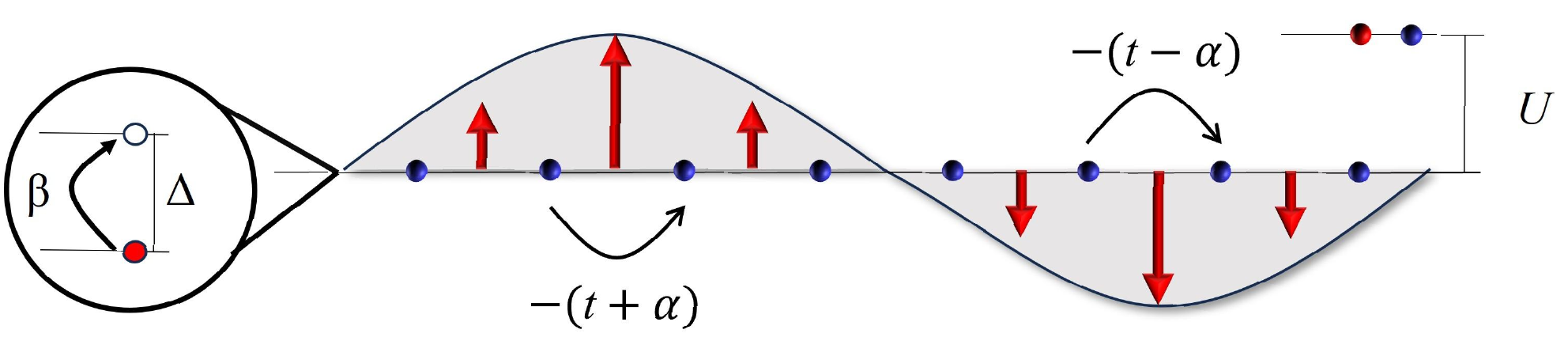}
		\caption{ Schematic illustration of the one-dimensional spin-1/2 fermion dynamical lattice model defined by the Hamiltonian (\ref{H1}). Itinerant fermions occupy the lattice sites and hop between nearest neighbors with amplitude $t$, which is dynamically modulated by localized spins through a coupling $\alpha$, leading to effective tunneling amplitudes $-(t \pm \alpha)$. The fermions are subject to a Zeeman field $h$, while the localized spins experience longitudinal $\Delta$ and transverse fields $\beta$. An onsite Hubbard interaction $U$ acts between opposite fermion. The inset illustrates the two-level structure of the localized spin degree of freedom.}
		\label{fig:1panel}
	\end{figure}

	The model we consider is a spin-1/2 fermion version of the 1D dynamical lattice, originally introduced in the context of bosonic analogs of the Peierls transition and the associated topological phenomena~\cite{2018prl}. In contrast to the conventional Ising–Kondo exchange, the spin-1/2 fermion dynamical lattice model considered here features a bond-centered dynamical coupling between itinerant fermions and localized spins. The motion of itinerant fermions is directly modulated by the local spin configuration through a coupling term that dynamically links the fermion hopping amplitude to the localized spin degrees of freedom. A schematic illustration of the model is shown in Fig.~\ref{fig:1panel}. The corresponding Hamiltonian can be expressed as ($\hbar=1$ throughout):
	\begin{eqnarray} \label{H1}
		\hat{H} &\!\!\!\!=\!\!\!\!& -t\! \sum_{i,\sigma}( \hat{c}_{i,\sigma}^{\dag} \hat{c}_{i+1,\sigma} +\!\mathrm{H.c.})-\alpha\sum_{i,\sigma}(\hat{c}_{i,\sigma}^{\dag}\hat{S}_{i}^{z}\hat{c}_{i+1,\sigma} +\!\mathrm{H.c.})  \notag \\
		&&+h\sum_{i}(\hat{n}_{i,\uparrow}-\hat{n}_{i,\downarrow})+ \frac{\Delta}{2}\sum_{i}\hat{S}_{i}^{z}+ \beta\sum_{i}\hat{S}_{i}^{x}\notag \\
		&&+U\sum_{i}\hat{n}_{i,\uparrow}\hat{n}_{i,\downarrow},
	\end{eqnarray}
	where $\hat{c}_{i,\sigma}^{\dag} $$(\hat{c}_{i,\sigma})$ is the creation (annihilation) field operator of the itinerant fermions with spin $\sigma (=\uparrow,\downarrow)$ at lattice site $i$ and $\hat{n}_{i,\sigma}=\hat{c}_{i,\sigma}^{\dag}\hat{c}_{i,\sigma}$ is the number operator. The itinerant fermions can hop between adjacent sites $i$, $i+1$ with the hopping rate $t$. $\hat{S}_{i}^{z}$ and $\hat{S}_{i}^{x}$ are spin-1/2 operators associated with a quantum spin residing on the bond between sites $i$ and $i+1$, representing respectively the longitudinal ($z$) and transverse ($x$) components of the localized spin degrees of freedom.
	
	With these definitions, the second term of Hamiltonian (1) describes a localized spin-dependent fermion hopping along each bond. The effective hopping amplitude across a bond is enhanced when the bond-center localized spin $\hat{S}_{i}^{z}$ is in the “up” state and suppressed when it is in the “down” state, thus directly linking the itinerant fermions' motion to the configuration of the bond-centered localized spins. And the third term acts as an effective Zeeman field applied to the itinerant fermions, introducing a spin-dependent energy shift between the two fermion spin components. By varying $h$, one can continuously tune the fermion population imbalance $N_{\uparrow}-N_{\downarrow}$, which effectively controls the spin polarization of the itinerant fermions and drives transitions between different magnetic phases. The next two terms describe the intrinsic spin dynamics of the bond-centered localized moments. The first one acts as an effective longitudinal field along the $z$ direction, favoring alignment of the localized spins, while the second introduces transverse quantum fluctuations that can induce localized spin flips and drive magnetic phase transitions. The direct repulsive interaction between itinerant fermions is included through the last Hubbard-$U$ term.
	This formulation naturally connects to $\mathbb{Z}_2$ gauge-coupled fermion systems~\cite{Z2} and quantum link models~\cite{qlm}, where bond-centered localized spins act as dynamical gauge fields mediating fermion motion~\cite{qlm,Z2_spinless}.
	
	In this work, we focus on the quasiadiabatic regime ($\beta \ll t$), setting the energy scale $t=1$ and $\beta=0.02$ unless otherwise specified. In this limit, the localized spin ground state is governed by the competition between the longitudinal field $\Delta$ and the spin-dependent hopping $\alpha$. A dominant $\Delta$ or $\alpha$ leads to a spatially uniform expectation value $\langle\hat{S}_{i}^{z}\rangle$. Conversely, when they are comparable, their competition induces nontrivial magnetic patterns, breaking translational symmetry and stabilizing modulated SDW orders.

	Here, we perform DMRG calculations to determine the many-body ground state of the Hamiltonian, from which various physical observables are evaluated. In our simulations the filling number $\rho=N/L$, with $N$ being the total number of fermions is conserved and can be tuned from 0 to 1. We typically consider lattice sizes of $L=24,36,48,60$ and impose open boundary conditions. For each lattice size, up to 600 states are kept per DMRG block and at least 60 sweeps are carried out, ensuring a maximum truncation error of order $10^{-9}$.
	\begin{figure}[t]
		\centering
		\includegraphics[width=0.5\textwidth]{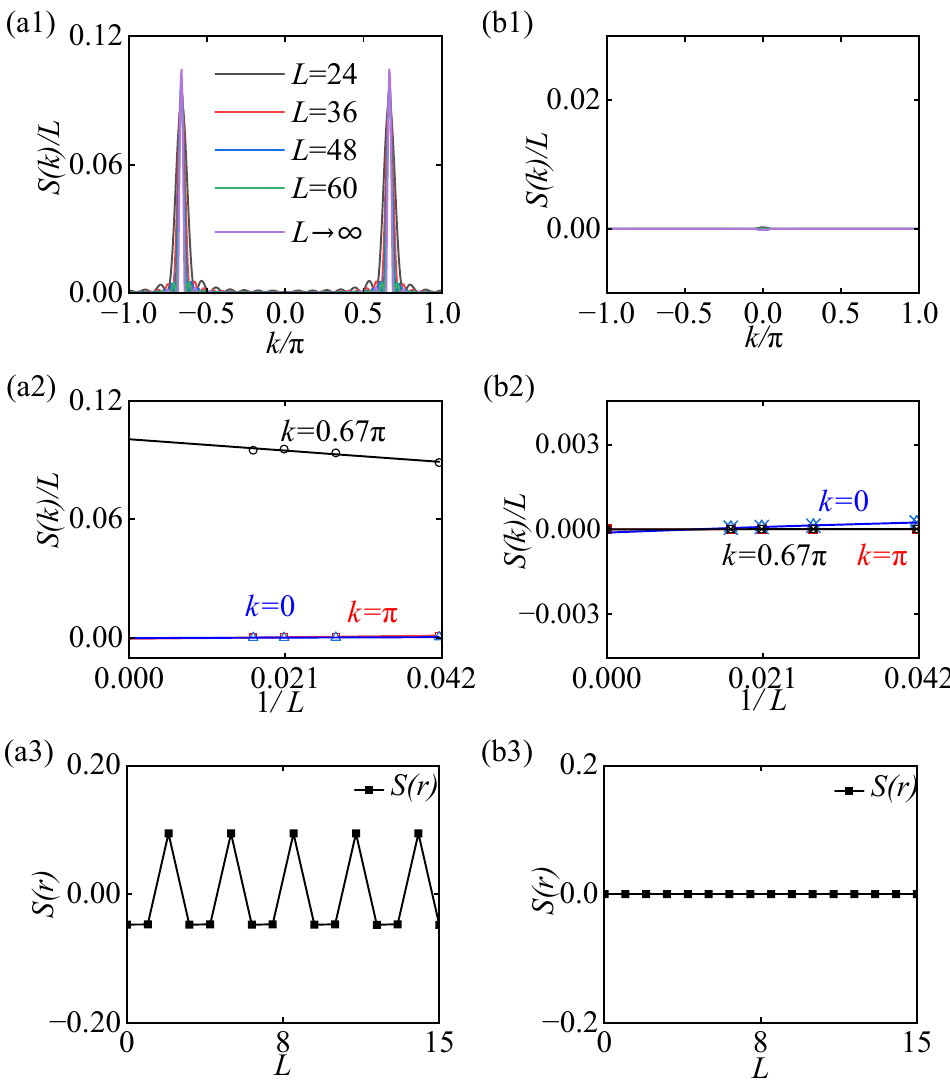}
		\caption{(a1) and (b1) The scaled spin structure factor $S^{z}(k)/L$, (a2) and (b2) the corresponding finite-size scalings at three characteristic wave vectors for systems. (a3) and (b3) The correlation function $S(r)$ for systems with $L = 30$. Different system sizes are characterized by lines with different colors. Panels (a1)–(a3) correspond to $\Delta=1.0$, while (b1)–(b3) correspond to $\Delta=0.1$ In these figures, we set $\rho=1$, $h=1.0$, $\alpha=0.5$, $\beta=0.02$ and $U=0$.}
		\label{fig:2panel}
	\end{figure}
	\section{RESULTS AND DISCUSSION}
	\subsection{Long-Range Order}
	Firstly, we demonstrate the ground-state in the absence of the Hubbard interaction (i.e., $U=0$). To systematically determine the magnetic properties, we evaluate the localized spin structure factor,
	\begin{equation} \label{sk}
		S^{z}(k) = \frac{1}{L} \sum_{i,j} e^{ik(i-j)} \langle (\hat{S}^{z}_{i}-\bar{S^{z}})(\hat{S}^{z}_{j}-\bar{S^{z}}) \rangle,
	\end{equation}
	with $\bar{S^{z}}=\sum_{i}\langle\hat{S}^{z}_{i}\rangle/L$, across different coupling regimes. The peak position $k^\mathrm{max}$ of $S^{z}(k)$ captures the dominant magnetic modulation of the localized spins. When the localized-spin-dependent hopping $\alpha$ and the longitudinal field $\Delta$ are comparable (for instance, setting $\alpha=0.5$ and $\Delta=1.0$), we observe the emergence of sharp peaks in the scaled structure factor $S^{z}(k)/L$ at finite wave vectors, as shown in Fig.~\ref{fig:2panel}(a1). Crucially, these peaks remain finite under finite-size scaling extrapolation, satisfying $\lim_{L \to \infty} S^{z}(k^\mathrm{max})/L > 0$~\cite{6}, as shown in Fig.~\ref{fig:2panel}(a2). Concurrently, when fixing the system size $L$, the corresponding real-space correlation function,
	\begin{equation} \label{sr}
		S^{z}(r) = \frac{1}{L} \sum_{l} \langle (\hat{S}^{z}_{l}-\bar{S^{z}})(\hat{S}^{z}_{l+r}-\bar{S^{z}}) \rangle,
	\end{equation}
	exhibits persistent oscillations without exponential decay at large distances $r$ [see Fig.~\ref{fig:2panel}(a3)]. These combined phenomena provide direct evidence of long-range magnetic modulation, allowing us to unambiguously identify this parameter regime as a stable SDW phase.
	
	In contrast, when one parameter strongly dominates over the other (for instance, setting $\alpha=0.5$ and $\Delta=0.1$, where $\alpha > \Delta$), the physical behavior changes fundamentally. The peak value of $S^{z}(k)/L$ shrinks and strictly vanishes in the thermodynamic limit, as demonstrated by the finite-size scaling, as shown in Figs.~\ref{fig:2panel}(b1) and~\ref{fig:2panel}(b2). Consistently, the corresponding real-space correlation function $S^{z}(r)$ shows no oscillatory behavior and instead decays rapidly [see Fig.~\ref{fig:2panel}(b3)]. This distinct lack of long-range magnetic correlations confirms that the system resides in a magnetically disordered PM phase.
	\begin{figure}[t]
		\centering
			\includegraphics[width=0.5\textwidth]{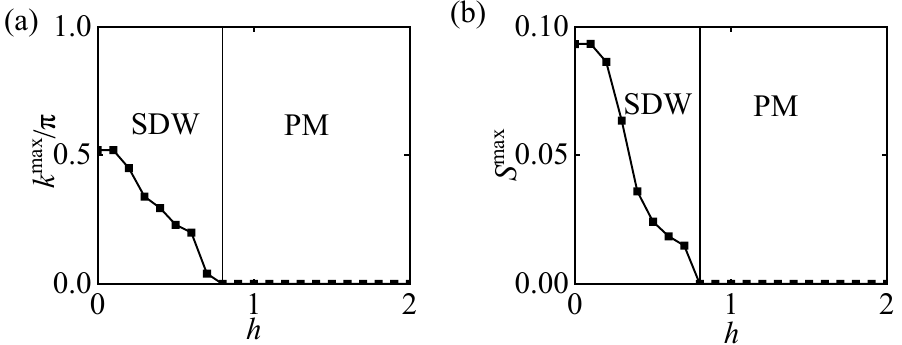}
		\caption{(a) The ordering wave vector $k^\mathrm{max}$ and (b) the ordering strength $S^\mathrm{max}$ as functions of $h$ with $\rho=1/2$, $\alpha=0.5$, $\Delta=1.0$, $\beta=0.02$ and $U=0$.}
		\label{fig:3panel}
    \end{figure}

	\subsection{Phase Transitions}
	To further characterize this magnetic phase transition, we drive the system with the external Zeeman field $h$ at filling $\rho=1/2$. As shown in Fig.~\ref{fig:3panel}(a), the ordering wave vector $k^\mathrm{max}$ decreases continuously from $\pi/2$ at $h=0$. This momentum shift reflects a gradual enlargement of the SDW modulation period as the fermion spin imbalance grows. To pinpoint the phase boundary, we monitor the thermodynamic ordering strength $S^\mathrm{max}=\lim_{L \to \infty} S^{z}(k^\mathrm{max})/L$ [see Fig.~\ref{fig:3panel}(b)]. $S^\mathrm{max}$ remains robust at small $h$, but decreases steadily until it strictly vanishes at a critical field $h_c \approx 0.7$, beyond which a well-defined $k^\mathrm{max}$ ceases to exist. The evolution of $k^\mathrm{max}$ and $S^\mathrm{max}$ clearly demonstrates a continuous quantum phase transition from the incommensurate SDW phase to the disordered PM state.
	
\begin{figure}[t]
		\centering
		\includegraphics[width=0.5\textwidth]{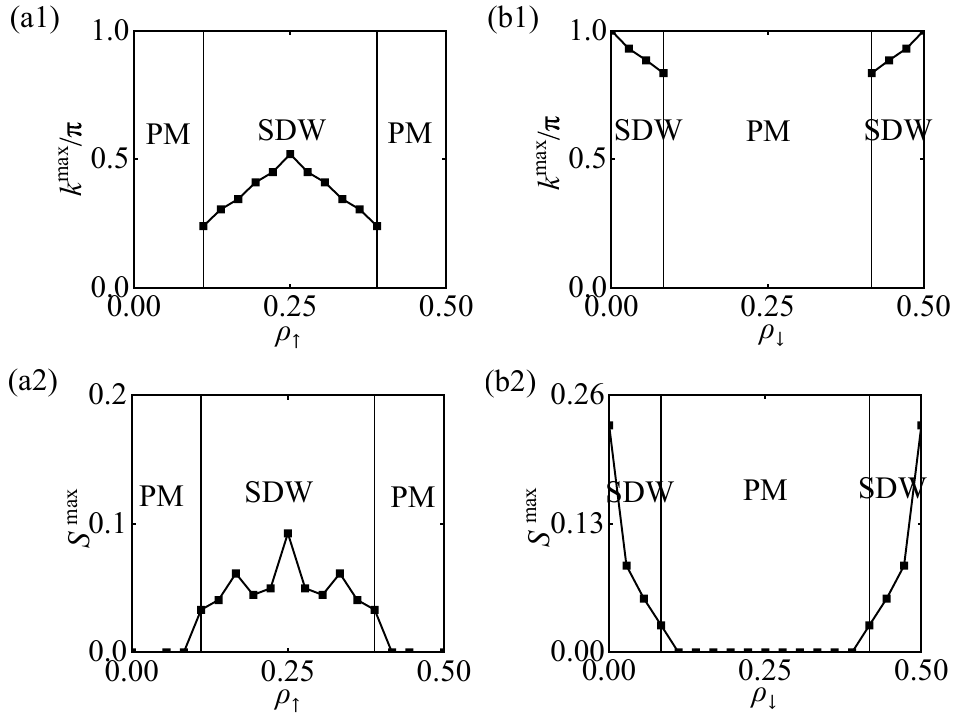}
		\caption{(a1) and (b1) The ordering wave vector $k^\mathrm{max}$  and (a2) and (b2) the ordering strength $S^\mathrm{max}$ as functions of $\rho_\sigma$ with (a1) and (a2) $\alpha=0.4$, (b1) and (b2) $\alpha=0.5$. In all panels, the parameters are fixed at $\rho=1/2$, $\Delta=0.65$, $\beta=0.02$ and $U=0$.}
		\label{fig:4panel}
	\end{figure}

	In 1D Hubbard-like models, Luttinger-liquid theory predicts that fermion population imbalance strongly modulates incommensurate spin correlations~\cite{15}. Consistently, our simulations reveal that the SDW wave vector is dictated by the density of the individual spin components, $\rho_{\sigma}$. As shown in Figs.~\ref{fig:4panel}(a1) and~\ref{fig:4panel}(a2), the system resides in the PM phase at both extreme limits of $\rho_{\uparrow}$ (nearly empty or highly polarized). However, in the intermediate partially-polarized regime, a robust SDW phase emerges. Here, the peak position $k^{\mathrm{max}}$ exhibits a strict linear dependence on the spin-up density $\rho_{\uparrow}$ with a positive slope. Conversely, as depicted in Figs.~\ref{fig:4panel}(b1) and ~\ref{fig:4panel}(b2) under different coupling parameters, the SDW phase stabilizes at both low and high densities of $\rho_{\downarrow}$, while the PM phase interrupts the intermediate region. In these phases, $k^{\mathrm{max}}$ instead follows a strict linear dependence on $\rho_{\downarrow}$ with a negative slope.

	In this case, the thermodynamic ordering strength $S^\mathrm{max}$ remains finite within a limited interval, clearly signaling the presence of robust long-range order. Depending on the specific parameter regime, the linear relationship between the dominant SDW ordering vector $k^\mathrm{max}$ and the spin densities ($\rho_{\uparrow}$, $\rho_{\downarrow}$) manifests in two distinct patterns. We denote the positive-slope pattern as SDW-$\mathrm{I}$ and the negative-slope pattern as SDW-$\mathrm{II}$, which follow the relations:
	\begin{equation} \label{dual-branch}
		\frac{k_{\max}}{\pi} \simeq
		\begin{cases}
			2\rho_\uparrow = 2\rho - 2\rho_\downarrow, & \text{SDW-I}, \\[6pt]
			2\rho_\downarrow = 2\rho - 2\rho_\uparrow, & \text{SDW-II}.
		\end{cases}
	\end{equation}
	These exact linear dependencies align with the predictions of Luttinger-liquid theory for incommensurate spin chains, where the ordering wave vector varies strictly with the fermion spin polarization~\cite{15}.
	
\begin{figure}[t]
		\centering
		\includegraphics[width=0.5\textwidth]{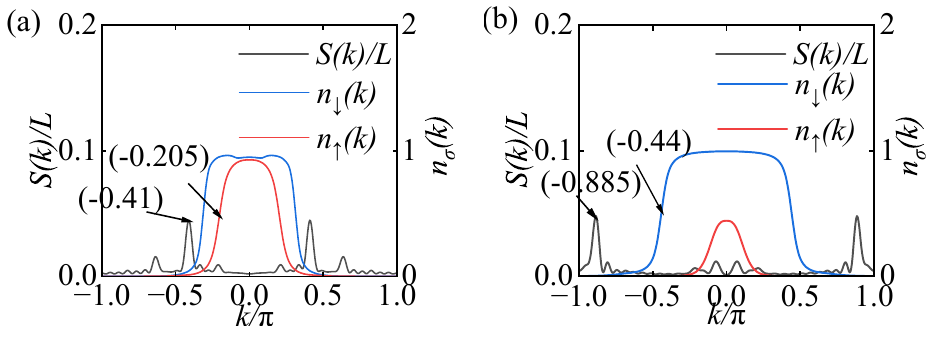}
		\caption{The momentum distribution function $n(k)$ and the spin structure factor $S^{z} (k)/L$ for systems with (a) $\alpha=0.4$ and $h=0.3$, (b) $\alpha=0.5$ and $h=0.8$. Different order are characterized by lines with different colors. In these figures, we set $L=36$, $\rho=1/2$, $\Delta=0.65$, $\beta=0.02$ and $U=0$.}
		\label{fig:5panel}
	\end{figure}

	To uncover the microscopic origin of these two SDW patterns, we evaluate the spin-resolved momentum distribution function of the itinerant fermions,
	\begin{equation} \label{nk}
		n_{\sigma}(k) = \frac{1}{L} \sum_{l,j} \langle c^{\dagger}_{l,\sigma} c_{j,\sigma} \rangle e^{i(l-j)k},
	\end{equation}
	with the total distribution given by $n(k)=n_{\uparrow} (k)+n_{\downarrow} (k)$. As illustrated in Fig.~\ref{fig:5panel}(a), in the SDW-$\mathrm{I}$ regime, the spin-up and spin-down Fermi surfaces are nearly symmetric, yielding an SDW peak at $k^\mathrm{max}/\pi \approx 2\rho_{\uparrow} = 2\rho - 2\rho_{\downarrow}$. This exactly matches the nesting vector connecting the spin-resolved Fermi points. In contrast, in the SDW-$\mathrm{II}$ regime [see Fig.~\ref{fig:5panel}(b)], the two fermion spin components develop distinct Fermi momenta, shifting the nesting vector and the corresponding SDW peak to $k^\mathrm{max}/\pi \approx 2\rho_{\downarrow} = 2\rho - 2\rho_{\uparrow}$.
	These results confirm that the SDW ordering vector is dictated not merely by the total filling $\rho$, but primarily by the relative occupations of the fermion spin components. From this viewpoint, the emergence of the dual-pattern SDW phase can be regarded as a magnetic analog of a Peierls-like transition driven by Fermi-surface nesting in the localized spin sector.
	
	\subsection{Phase Diagrams}
	\begin{figure}[t]
		\centering
		\includegraphics[width=0.5\textwidth]{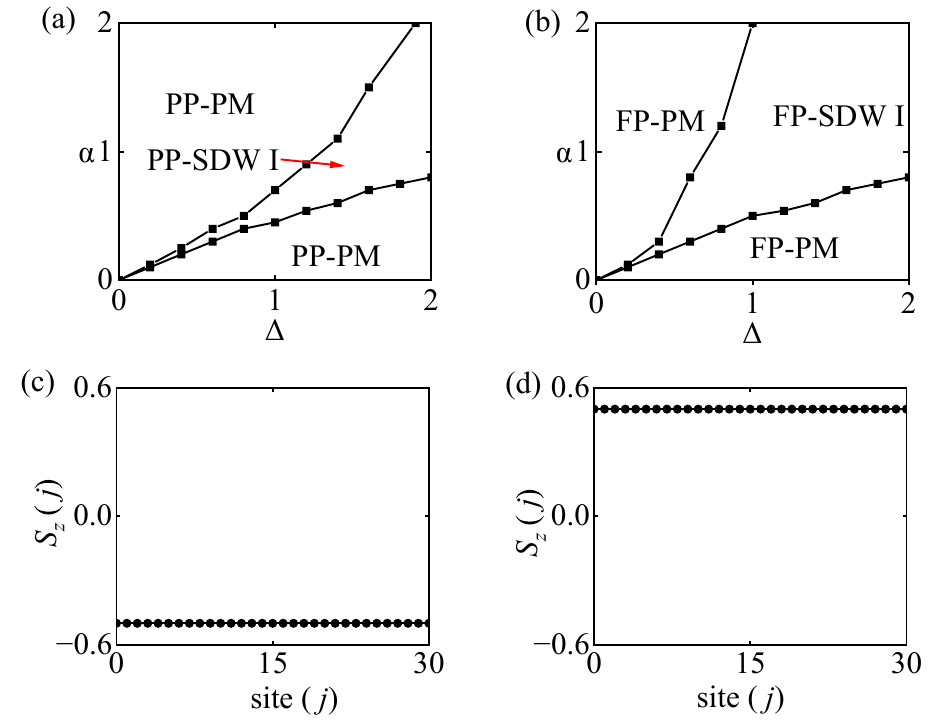}
		\caption{The phase diagram in the $\alpha-\Delta$ plane with (a) $h = 0.5$ (b) $h = 2.0$. The phase boundaries in (a)-(b) have been extrapolated to the thermodynamic limit $L \to \infty$. The localized spin structure expectation values with $L=30$ and $h=0.5$, for (c) $\alpha$ = 0.1, $\Delta = 1.2$ ($\Delta\gg\alpha$), (d) $\alpha$ = 1.2, $\Delta=0.1$ ($\Delta\ll\alpha$). In these figures, we set $\rho =1/2$, $\beta = 0.02$ and $U = 0$.}
		\label{fig:6panel}
	\end{figure}
	Having established the microscopic mechanism of the SDW modulations, we now map out the global phase boundaries. Figures~\ref{fig:6panel}(a) and~\ref{fig:6panel}(b) present the phase diagrams in the $\alpha-\Delta$ plane for the partially polarized (PP) fermion regime with $h=0.5$, and fully polarized (FP) fermion regime with $h=2.0$. The coupling parameters play complementary yet competing roles: the kinetic hopping $\alpha$ promotes itinerant-localized correlations and stabilizes the SDW order, whereas the longitudinal field $\Delta$ favors a uniform spin alignment along the $z$-axis, suppressing transverse fluctuations. When $\alpha$ and $\Delta$ are comparable, their competition induces spatially modulated localized spin configurations. Conversely, when one term dominates ($\Delta\gg\alpha$ or $\Delta\ll\alpha$), the localized spins uniformly polarize to $\langle\hat{S}_{i}^{z}\rangle \approx -1/2$ or $+1/2$, driving the system into a PM state [see Figs.~\ref{fig:6panel}(c) and ~\ref{fig:6panel}(d)]. Notably, enhancing both $\alpha$ and $\Delta$ simultaneously expands the SDW region, confirming that this magnetic order stems fundamentally from their delicate balance, while $h$ primarily tunes the overall fermion polarization.
	
	\begin{figure}[t]
		\centering
		\includegraphics[width=0.5\textwidth]{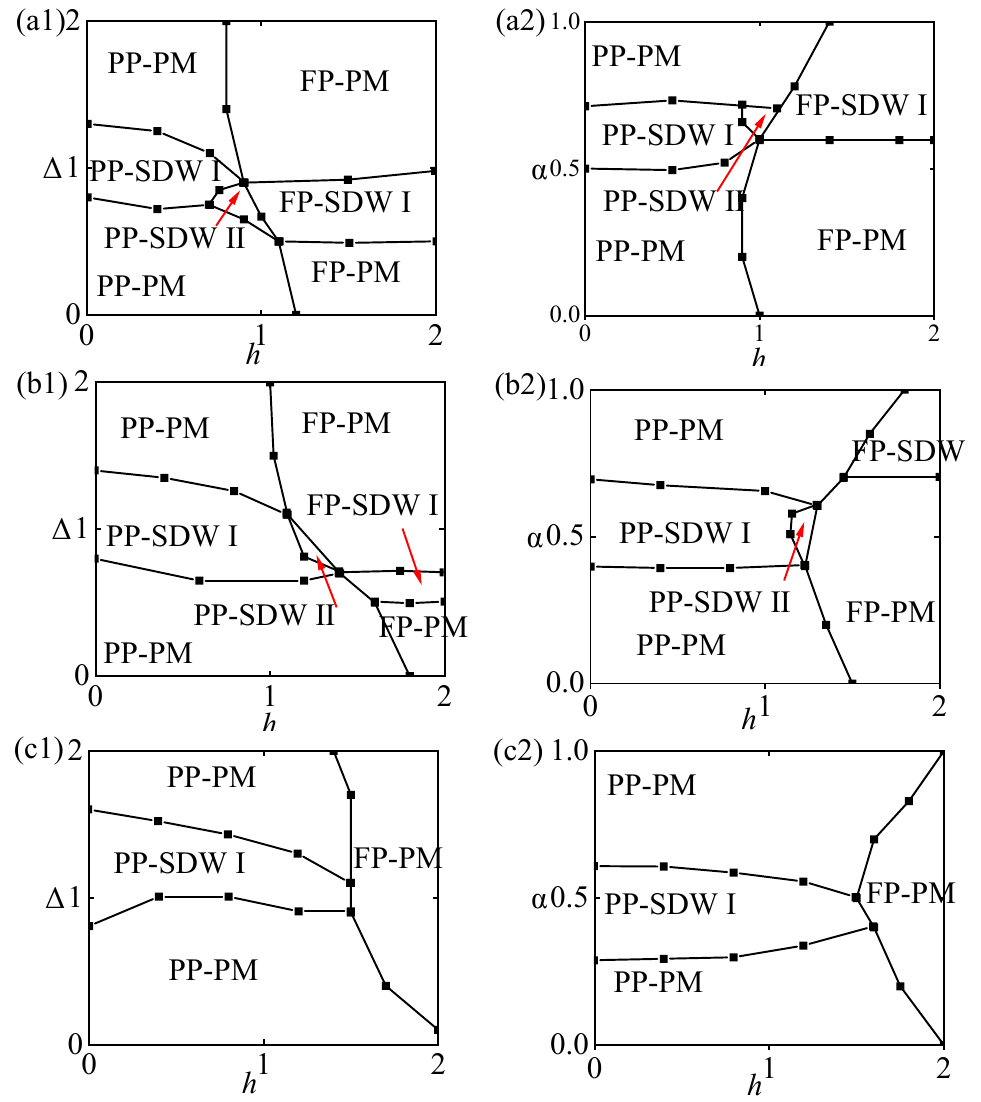}
		\caption{The phase diagrams in the $h-\Delta$ (left column) with $\alpha=0.5$, and $h-\alpha$ (right column) with $\Delta=1.0$ planes at different fillings $\rho=1/2$, $2/3$ and $1$. The phase boundaries have been extrapolated to the thermodynamic limit $L \to \infty$. In these figures, we set $\beta=0.02$ and $U=0$.}
		\label{fig:7panel}
	\end{figure}
	To further quantify this interplay, we summarize the phase diagrams in the $h-\Delta$ and $h-\alpha$ planes across different fillings ($\rho=1/2$, $2/3$, and $1$), as shown in Fig.~\ref{fig:7panel}. In the PP regime at generic fractional fillings ($\rho<1$), tuning $\alpha$ and $\Delta$ can stabilize both SDW-I and SDW-II patterns. However, as the system approaches half-filling ($\rho=1$), the SDW-II region shrinks and vanishes, leaving only the SDW-I pattern. This indicates a strong preference for symmetric localized spin modulation near half-filling. In the FP regime, the system effectively reduces to a single-component fermion gas, and the SDW ordering strictly follows the Luttinger-liquid prediction $k^\mathrm{max}=2k_{F}=2\pi\rho$. Overall, the stability and specific modulation pattern of the SDW are cooperatively controlled by the kinetic coupling $\alpha$, the longitudinal field $\Delta$, the external Zeeman field $h$, and the filling factor $\rho$.

    \section{EFFECT OF THE REPULSIVE HUBBARD INTERACTION}
\begin{figure}[b]
	\includegraphics[width=0.5\textwidth]{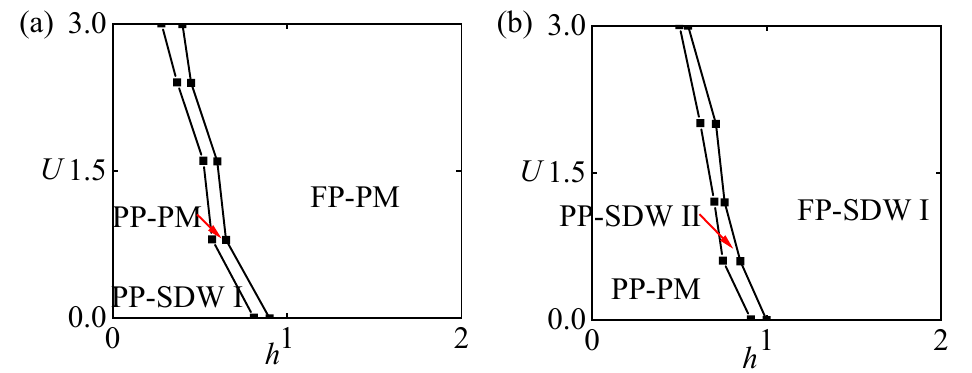}
	\caption{The phase diagram in the $h-U$ plane at $\rho=1/2$, $\Delta=0.65$, and $\beta=0.02$ for different coupling strengths $\alpha$ with (a) $\alpha=0.4$ and (b) $\alpha=0.5$.}
	\label{fig:8panel}
    \end{figure}

    We now turn to the influence of the onsite repulsive Hubbard interaction $U$ on the magnetic phase structure. Figure~\ref{fig:8panel} presents the $h-U$ phase diagrams for the SDW-I and SDW-II phases under varying $h$. The results show that the inclusion of a finite $U$ does not qualitatively alter the nature of SDW-I and SDW-II phases identified previously. However, the repulsive interaction significantly affects the magnetic polarization tendency of the system. As $U$ increases, the system becomes more susceptible to fermion spin polarization, and the critical field required to reach the fermion spin FP phase is reduced. This behavior can be attributed to the suppression of double occupancy by the repulsive interaction, which enhances the sensitivity of the remaining fermion spin degrees of freedom to the external Zeeman field. Consequently, $U$ primarily affects the fermion spin polarization phase, without qualitatively modifying the nature of the SDW order.
    \section{CONCLUSIONS}
    In summary, we have investigated the ground-state magnetic properties of the 1D spin-1/2 fermion dynamical lattice model using the DMRG method. Our results reveal two distinct quantum phases: the PM phase and the SDW phase. The SDW phase further exhibits two characteristic phenomena, SDW-I and SDW-II, depending on the interplay between the coupling parameters. Each pattern of the SDW is characterized by a well-defined ordering wave vector, which coincides with the nesting wave vector of the Fermi surfaces of spin-up and spin-down itinerant fermions, respectively. This correspondence implies that the PM–SDW transition in the dynamical lattice can be viewed as a magnetic analog of the Peierls transition, which occurs in the charge sector of 1D metals. Furthermore, the influence of the repulsive Hubbard interaction between itinerant fermions has also been clarified: while it enhances the fermion spin polarization tendency, it does not alter the fundamental nature of the SDW-I and SDW-II phases. Altogether, this study provides a new paradigm for controlling magnetic modulations and enriches our fundamental understanding of quantum phase transitions in dynamical lattice many-body systems.
	\section*{ACKNOWLEDGMENTS}
	We thank Jingtao Fan for helpful discussions and Miles Stoudenmire for assistance with the code. This work is supported by the National Key R\&D Program of China under Grant No.~2022YFA1404201, the National Natural Science Foundation of China (NSFC) under Grant No.~12574297, Fundamental Research Program of Shanxi Province under Grant No.~202403021221024, the Research Project Supported by Shanxi Scholarship Council of China and Shanxi Grant No.~1331KSC. Our simulations use the ITensor library~\cite{ITensor}.


\begin{thebibliography}{99}
	
	\bibitem{1} E. Dagotto, Correlated fermions in high-temperature superconductors, Rev. Mod. Phys. \textbf{66}, 763 (1994).
	
	\bibitem{2} M. Imada, A. Fujimori, and Y. Tokura, Metal-insulator transitions, Rev. Mod. Phys. \textbf{70}, 1039 (1998).
	
	\bibitem{3} P. A. Lee, N. Nagaosa, and X.-G. Wen, Doping a Mott insulator: Physics of high-temperature superconductivity, Rev. Mod. Phys. \textbf{78}, 17 (2006).
	
	\bibitem{4} T. Moriya, \textit{Spin Fluctuations in Itinerant Electron Magnetism} (Springer, Berlin, 1985).
	
	\bibitem{5} P. W. Anderson, \textit{Basic Notions of Condensed Matter Physics} (CRC Press, Boca Raton, 1997).
	
	\bibitem{6} A. Auerbach, \textit{Interacting Electrons and Quantum Magnetism} (Springer, New York, 1994).
	
	\bibitem{7} S. Sachdev, \textit{Quantum Phase Transitions} (Cambridge University Press, Cambridge, 2011).
	
	\bibitem{8} L. Balents, Spin liquids in frustrated magnets, Nature \textbf{464}, 199 (2010).
	
	\bibitem{9} L. Savary and L. Balents, Quantum spin liquids: a review, Rep. Prog. Phys. \textbf{80}, 016502 (2017).
	
	\bibitem{10} Q.-M. Si and F. Steglich, Heavy fermions and quantum phase transitions, Science \textbf{329}, 1161 (2010).
	
	\bibitem{11} M. Vojta, Quantum phase transitions, Rep. Prog. Phys. \textbf{66}, 2069 (2003).
	
	\bibitem{12} X.-G. Wen, \textit{Quantum Field Theory of Many-Body Systems} (Oxford University Press, Oxford, 2007).
	
	\bibitem{13} B. Keimer, S. A. Kivelson, M. R. Norman, S. Uchida, and J. Zaanen, From quantum matter to high-temperature superconductivity in copper oxides, Nature \textbf{518}, 179 (2015).
	
	\bibitem{DagottoRice1996} E. Dagotto and T. M. Rice, Surprises on the way from one- to two-dimensional quantum magnets, Science \textbf{271}, 618 (1996).
	
	\bibitem{14} M. D. Johannes and I. I. Mazin, Fermi surface nesting and the origin of charge density waves in metals, Phys. Rev. B \textbf{77}, 165135 (2008).
	
	\bibitem{16} M. Fabrizio, Role of transverse hopping in a two-coupled-chains model, Phys. Rev. B \textbf{48}, 15838 (1993).
	
	\bibitem{super1} P. Fulde and R. A. Ferrell, Superconductivity in a Strong Spin-Exchange Field, Phys. Rev. \textbf{135}, A550 (1964).

    \bibitem{super2} A. I. Larkin and Y. N. Ovchinnikov, Nonuniform state of superconductors, Sov. Phys. JETP \textbf{20}, 762 (1964).	

	\bibitem{17} X.-F. Zhou, J.-T. Fan, and S.-T. Jia, Magnetic order and strongly correlated effects in the one-dimensional Ising-Kondo lattice, Phys. Rev. B \textbf{109}, 195112 (2024).
	
	\bibitem{18} F. H. L. Essler, H. Frahm, F. G\"ohmann, A. Kl\"umper, and V. E. Korepin, \textit{The One-Dimensional Hubbard Model} (Cambridge University Press, Cambridge, 2005).
	
	\bibitem{23} H. J. Schulz, Correlation exponents and the metal-insulator transition in the one-dimensional Hubbard model, Phys. Rev. Lett. \textbf{64}, 2831 (1990).
	
	\bibitem{24} J. Voit, One-dimensional Fermi liquids, Rep. Prog. Phys. \textbf{58}, 977 (1995).
	
	\bibitem{jiexi} M. Ogata and H. Shiba, Bethe-ansatz wave function, momentum distribution, and spin correlation in the one-dimensional strongly correlated Hubbard model, Phys. Rev. B \textbf{41}, 2326 (1990).
	
	\bibitem{15} T. Giamarchi, \textit{Quantum Physics in One Dimension} (Oxford University Press, Oxford, 2003).
	
	\bibitem{21} M. Nakagawa and N. Kawakami, Laser-induced Kondo effect in ultracold alkaline-earth fermions, Phys. Rev. Lett. \textbf{115}, 165303 (2015).
	
	\bibitem{22} S. Hoshino, J. Otsuki, and Y. Kuramoto, Itinerant antiferromagnetism in infinite dimensional Kondo lattice, Phys. Rev. B \textbf{81}, 113108 (2010).
	
	\bibitem{25} C. Kim \textit{et al.}, Observation of spin-charge separation in one-dimensional SrCuO$_2$, Phys. Rev. Lett. \textbf{77}, 4054 (1996).
	
	\bibitem{26} B. Lake, D. A. Tennant, C. D. Frost, and S. E. Nagler, Quantum criticality and universal scaling of a quantum antiferromagnet, Nat. Mater. \textbf{4}, 329 (2005).
	
	\bibitem{27} S. Tomonaga, Remarks on Bloch's Method of Sound Waves applied to Many-Fermion Problems, Prog. Theor. Phys. \textbf{5}, 544 (1950).
	
	\bibitem{28} J. M. Luttinger, An Exactly Soluble Model of a Many-Fermion System, J. Math. Phys. \textbf{4}, 1154 (1963).
	
	\bibitem{29} F. D. M. Haldane, ``Luttinger liquid theory'' of one-dimensional quantum fluids. I. Properties of the Luttinger model and their extension to the general 1D interacting spinless Fermi gas, J. Phys. C: Solid State Phys. \textbf{14}, 2585 (1981).
	
    \bibitem{20} E. Jeckelmann, Dynamical density-matrix renormalization-group method, Phys. Rev. B \textbf{66}, 045114 (2002).

   	\bibitem{41} A. Dhar, J. J. Kinnunen, and P. T\"orm\"a, Population imbalance in the extended Fermi-Hubbard model, Phys. Rev. B \textbf{94}, 075116 (2016).

    \bibitem{qmc1} R. Blankenbecler, D. J. Scalapino, and R. L. Sugar, Monte Carlo calculations of coupled boson-fermion systems, Phys. Rev. D \textbf{24}, 2278 (1981).

    \bibitem{qmc2} S. R. White, D. J. Scalapino, R. L. Sugar, E. Y. Loh, J. E. Gubernatis, and R. T. Scalettar, Numerical study of the two-dimensional Hubbard model, Phys. Rev. B \textbf{40}, 506 (1989).
    \bibitem{32} K. Yang, Inhomogeneous superconducting state in quasi-one-dimensional systems, Phys. Rev. B \textbf{63}, 140511(R) (2001).

    \bibitem{33} L. Radzihovsky and D. E. Sheehy, Imbalanced Feshbach-resonant Fermi gases, Rep. Prog. Phys. \textbf{73}, 076501 (2010).

    \bibitem{39} M. Okumura, S. Yamada, M. Machida, and H. Aoki, Phase-separated ferromagnetism in a spin-imbalanced system of Fermi atoms loaded in an optical ladder: A density-matrix renormalization-group study, Phys. Rev. A \textbf{83}, 031606(R) (2011).

    \bibitem{35} A. Korolyuk, F. Massel, and P. T\"orm\"a, Probing the Fulde-Ferrell-Larkin-Ovchinnikov phase by Double Occupancy Modulation Spectroscopy, Phys. Rev. Lett. \textbf{104}, 236402 (2010).

    \bibitem{37} G. Orso, Attractive Fermi Gases with Unequal Spin Populations in Highly Elongated Traps, Phys. Rev. Lett. \textbf{98}, 070402 (2007).

    \bibitem{38} H. Mosadeq and R. Asgari, Quantum phases of a one-dimensional dipolar Fermi gas, Phys. Rev. B \textbf{91}, 085126 (2015).

    \bibitem{40} Y.-A. Liao \textit{et al.}, Spin-imbalance in a one-dimensional Fermi gas, Nature \textbf{467}, 567 (2010).

    \bibitem{42} R. M. Lutchyn, M. Dzero, and V. M. Yakovenko, Spectroscopy of the soliton lattice formation in quasi-one-dimensional fermionic superfluids with population imbalance, Phys. Rev. A \textbf{84}, 033609 (2011).

    \bibitem{dmrg1} S. R. White, Density matrix formulation for quantum renormalization groups, Phys. Rev. Lett. \textbf{69}, 2863 (1992).

    \bibitem{dmrg2} U. Schollw\"ock, The density-matrix renormalization group, Rev. Mod. Phys. \textbf{77}, 259 (2005).

    \bibitem{ITensor} M. Fishman, S. R. White, and E. M. Stoudenmire, The ITensor software library for tensor network calculations, SciPost Phys. Codebases \textbf{4} (2022).

	\bibitem{2018prl} D. G. Cuadra, P. R. Grzybowski, A. Dauphin, and M. Lewenstein, Strongly Correlated Bosons on a Dynamical Lattice, Phys. Rev. Lett. \textbf{121}, 090402 (2018).
	
    \bibitem{Z2} A. Das, U. Borla, and S. Moroz, Fractionalized holes in one-dimensional $Z_{2}$ gauge theory coupled to fermion matter: Deconfined dynamics and emergent integrability, Phys. Rev. B \textbf{107}, 064302 (2023).

	\bibitem{qlm} U. J. Wiese, Ultracold quantum gases and lattice systems: quantum simulation of lattice gauge theories, Ann. Phys. (Berlin) \textbf{525}, 777--796 (2013).
	
	\bibitem{Z2_spinless} U. Borla, R. Verresen, F. Grusdt, and S. Moroz, Confined Phases of One-Dimensional Spinless Fermions Coupled to $Z_2$ Gauge Theory, Phys. Rev. Lett. \textbf{124}, 120503 (2020).
\end{thebibliography}
\end{document}